\documentclass[12pt]{article}
\usepackage{pdproc,epsfig} 

\begin{document}

\renewcommand{\thefootnote}{\alph{footnote}}
  
\title{
PROBING NON-STANDARD NEUTRINO PHYSICS AT \\
NEUTRINO FACTORY AND T2KK\protect\footnote{
Written version of a talk presented at the 
``Fourth International Workshop on Neutrino Oscillations in Venice'' 
(NO-VE 2008), Venice, Italy, 15-18, April 2008.}} 

\author{HISAKAZU MINAKATA}

\address{Department of Physics, Tokyo Metropolitan University, \\
1-1 Minami-Osawa, Hachioji, Tokyo 192-0397, Japan\\
 {\rm E-mail: minakata@phys.metro-u.ac.jp}}




\abstract{
We discuss ways to explore non-standard interactions (NSI) which 
neutrinos may possess by expressing them as effective four Fermi 
operators with coefficient of the order of 
$(M_{W} / M_{NP})^2 \sim10^{-2}$($10^{-4}$)  
for energy scales of new physics as $M_{NP}\sim1$(10) TeV.  
Neutrino Factory is a prime candidate for such apparatus 
that can reach to the extreme precision. 
I describe a two detector setting, one at baseline $L\sim3000$ km 
and the other at $L\sim7000$ km, which is able to solve the notorious 
$\theta_{13}-$NSI confusion, and possibly also the two-phase confusion. 
The resultant sensitivities to off-diagonal NSI elements 
$\varepsilon$'s are excellent, 
$|\varepsilon_{e \tau}| \simeq \mbox{a few} \times10^{-3}$ and 
$|\varepsilon_{e \mu}| \simeq \mbox{a few} \times10^{-4}$. 
Our results suggest a new picture of neutrino factory as a hunting 
machine for NSI while keeping its potential of precision measurement 
of lepton mixing parameters. 
Sensitivities to NSI by T2KK and the related settings are also discussed. 
}
   
\normalsize\baselineskip=15pt

\section{Introduction}

This conference is sub-titled as 
``Ten Years after the Neutrino Oscillations''. 
It refers an unforgettable event which occurred in Neutrino 1998 
conference in Takayama, Japan. 
The presentation by Kajita-san of atmospheric neutrino observation by 
Super-Kamiokande group \cite{kajita98} gave the first evidence for 
neutrino oscillation \cite{SKatm-evidence}, 
which received a long lasting ovation. 
But, as Koshiba-san pointed out in his presentation \cite{koshiba}, 
there was a prehistory to that event. 
The Kamiokande II experiment\footnote{
In 1986 the Kamiokande detector started its phase II operation 
armed with lowered energy threshold to observe solar neutrinos, 
which soon blossomed as neutrino detection from SN1987A 
\cite{SN1987a}. 
}
reported the deficit of muon-like events in its atmospheric neutrino 
observation in 1988 \cite{Kam-deficit}, 
the anomaly in ratio of muon-type to electron-type 
neutrino events  in 1992 \cite{Kam-ratio}, 
and then the anomalous zenith angle dependence of muon-type 
events in 1994 \cite {Kam-zenith}. 
In particular, the latter is strongly indicative of neutrino oscillation. 
The prehistory is reflected by the fact that the speaker in Neutrino 1998 represented not only Super-Kamiokande group but also Kamiokande II 
collaboration, as recollected in my slides in this conference 
\cite{NO-VE08-mina}. 
The anomaly was confirmed unambiguously by the high-statistics 
observation by Super-Kamiokande experiment. 
In the context of three-flavor neutrino mixing, this establishes 
neutrino oscillation \cite{SKatm-oscillation} in the 2-3 sector of 
the MNS matrix \cite{MNS}.

Ten years from Takayama declaration, as everybody knows, 
has been full of excitement. 
The solar neutrino experiment \cite{solar}, 
which was pioneered by Ray Davis 40 years ago \cite{davis}, 
finally wrote its conclusion that the cause of the 
solar neutrino problem is not due to our ignorance of interior of the sun 
but to neutrino flavor transformation \cite{solar-conclusion}. 
The KamLAND reactor neutrino experiment \cite{KamLAND} 
gave the first proof that neutrino oscillation takes place also in 
the 1-2 sector of the MNS matrix with parameters appropriate 
for the solar neutrino deficit. 
By excluding various other mechanisms of neutrino flavor transformation, 
it solved the solar neutrino problem. 
It is impressive to see evidence for spectral distortion of reactor 
antineutrinos at more than $5\sigma$ \cite{KamLAND_new}. 
The evidence for atmospheric neutrino oscillation was followed by 
confirmation by the accelerator neutrino experiments, 
one in Japan \cite{K2K} and the other in US \cite{MINOS}. 
Now, everybody agrees that neutrinos have masses and they oscillate.

\section{A Bold Question}

The important goal of the next generation accelerator \cite{T2K,NOVA}  
and the reactor \cite{reactor-proposal,reactor-exp} neutrino 
experiments is to measure $\theta_{13}$.
Fortunately, rich programs exist to serve for this purpose. 
If $\theta_{13}$ is large enough we may be able to proceed 
to search for leptonic CP violation. 
If the experiments have sufficient sensitivities to the matter effect, 
they may be able to determine the neutrino mass hierarchy. 

Suppose in some day all these goals are met and the MNS 
matrix elements are measured with precision comparable to 
those of CKM matrix \cite{cabibbo,KM}. 
Then, one might ask;
``Is this the final goal of neutrino experiments?''
I argue that the answer is {\em NO}. 
Of course, my argument cannot be a solid one. 
Let me, however,  mention it anyway.

\begin{itemize}

\item
Neutrinos are proved to be useful probe into physics beyond the 
Standard Model. Why should we believe that it is merely an accident?

\item
Cosmological neutrinos will soon become one of our machineries for 
probing nature \cite{halzen}. It is natural to suspect that they will bring us 
something entirely new. 

\item

People already suspected several candidates; 
Non-standard interactions, quantum decoherence, 
Lorentz-invariance violation, etc.

\end{itemize}

\noindent
In this talk, I concentrate on non-standard interactions (NSI) 
\cite{wolfenstein,valle,guzzo} which might be possessed by 
neutrinos.\footnote{
Of course, I do not say that the items above complete the all that 
should be in the list.
For example, Majorana nature of neutrino must be demonstrated, 
so important to understand leptogenesis \cite{leptogenesis},  
for example, 
as emphasized by Yoshimura-san in his talk \cite{yoshimura}. 
}
My presentation will be based on the two references 
\cite {NSI-nufact,NSP-T2KK}. 
There exist numerous references which devoted to this topics. 
Therefore, I would like to apologize, before start, to those who 
are not mentioned in my reference. More bibliography is 
contained in these papers.

\section{Non-Standard Interactions of Neutrinos}

Suppose that there is a new physics at energy scale greater than 
$\sim 1$ TeV. I denote the energy scale as $M_{NP}$. 
Then, it is natural to expect that higher-dimentional operators 
would exist which gives rise to effective new interactions of 
neutrinos with matter \cite{grossmann,berezhiani} 
\begin{eqnarray}
{\cal L}_{\mbox{eff}}^{\mbox{NSI}} = 
-2\sqrt{2}\, \varepsilon_{\alpha\beta}^{fP} G_F
(\overline{\nu}_\alpha \gamma_\mu P_L \nu_\beta)\,
(\overline{f} \gamma^\mu P f),
\label{LNSI}
\end{eqnarray}
where $G_F$ is the Fermi constant, and 
$f$ stands for the index running over fermion species in the earth, 
$f = e, u, d$, in which we follow \cite{davidson} for notation.\footnote{
There remains a serious question of whether effective dimension 
six operators like (\ref{LNSI}) which are consistent with severe 
constraints on charged lepton counterpart which is related by 
SU(2) gauge rotation. This point which was first addressed in 
\cite{berezhiani} 
is emphasized to me by Belen Gavela \cite{gavela-private}. 
}
$P$ stands for a projection operator and is either
$P_L\equiv \frac{1}{2} (1-\gamma_5)$ or 
$P_R\equiv \frac{1}{2} (1+\gamma_5)$. 

To summarize its effects on neutrino propagation it is customary 
to introduce the $\varepsilon$ parameters, which are defined as
$\varepsilon_{\alpha\beta} \equiv \sum_{f,P} \frac{n_f}{n_e}
\varepsilon_{\alpha\beta}^{fP}$, 
where $n_f$ is the number density of the fermion species $f$ in matter. 
Approximately, the relation 
$\varepsilon_{\alpha\beta} \simeq \sum_{P}
\left(
\varepsilon_{\alpha\beta}^{eP}
+ 3 \, \varepsilon_{\alpha\beta}^{uP}
+ 3 \, \varepsilon_{\alpha\beta}^{dP}
\right)$  
holds because of a factor of $\simeq$3 larger number of 
$u$ and $d$ quarks than electrons in iso-singlet matter. 
Using the $\varepsilon$ parameters the neutrino evolution equation 
which governs the neutrino propagation in matter is given as  
\begin{eqnarray} 
i {d\over dt} \left( \begin{array}{c} 
                   \nu_e \\ \nu_\mu \\ \nu_\tau 
                   \end{array}  \right)
 = \frac{1}{2E} \left[ U \left( \begin{array}{ccc}
                   0   & 0          & 0   \\
                   0   & \Delta m^2_{21}  & 0  \\
                   0   & 0           &  \Delta m^2_{31}  
                   \end{array} \right) U^{\dagger} +  
                  a \left( \begin{array}{ccc}
            1 + \varepsilon_{ee}     & \varepsilon_{e\mu} & \varepsilon_{e\tau} \\
            \varepsilon_{e \mu }^*  & \varepsilon_{\mu\mu}  & \varepsilon_{\mu\tau} \\
            \varepsilon_{e \tau}^* & \varepsilon_{\mu \tau }^* & \varepsilon_{\tau\tau} 
                   \end{array} 
                   \right) \right] ~
\left( \begin{array}{c} 
                   \nu_e \\ \nu_\mu \\ \nu_\tau 
                   \end{array}  \right)
\label{general-evolution}
\end{eqnarray}
where $U$ is the MNS matrix, and
$a\equiv 2 \sqrt 2 G_F n_e E$ \cite{wolfenstein} where $E$ is the
neutrino energy and $n_e$ denotes the electron number density along
the neutrino trajectory in the earth.  $\Delta m^2_{i j} \equiv
m^2_{i} - m^2_{j}$ with neutrino mass $m_{i}$ ($i=1-3$).
The phase of $\varepsilon$ parameters may provide new 
sources of CP violation \cite{concha1}.

NSI comes in not only into neutrino propagation but also to 
neutrino production and detection processes \cite{grossmann}.  
The current bounds on $\varepsilon_{\alpha\beta}^{fP}$ 
are obtained at 90\% CL \cite{davidson}  and at 95\% CL \cite{lep}. 
When translated (in a bold way!) into the $\varepsilon$ parameters 
defined above they may read as follows \cite{KSY06}: 
\begin{eqnarray}
\left[
\begin{array}{ccc}
-4 < \varepsilon_{ee} < 2.6 & |\varepsilon_{e \mu}| < 3.8 \times 10^{-4} & |\varepsilon_{e \tau}| < 1.9 \\
 & -0.05 < \varepsilon_{\mu \mu} < 0.08 &  | \varepsilon_{\mu \tau} | <  0.25  \\
  &  & | \varepsilon_{\tau \tau} | < 18.6  \\
\end{array}
\right]. 
\label{bound}
\end{eqnarray}
I emphasize that it is important to constrain the NSI parameters 
by various experiments. The bound placed by 
the atmospheric \cite{concha2,fornengo,friedland} and 
the solar neutrino experiments \cite{NSI-solar} are extensively discussed. 
It is also proposed that several low energy neutrino
experiments may be able to place equally severe constraints on NSI
\cite{barranco,scholberg,bueno}.  The bounds from them are placed on
the product of NSI at the source and the detection.

In this talk I concentrate on hunting NSI parameters during 
neutrino propagation. 
It is the part that can be dealt with in a model-independent 
manner and free from the ``unitarity violation''. 
By contrast, the way NSI comes in into production and detection 
processes is model-dependent.\footnote{
It appears to me that the main difference between our and 
the ``unitarity violation'' approach \cite{gavela} exists in that 
the latter chooses to specify a model (or a class of models) to 
allow them to relate the propagation $\varepsilon$'s to 
the production and the detector $\varepsilon$'s. 
}
Therefore, categorizing the model predictions is necessary 
before taking them into account. 
Moreover, I call the readers' attention to the fact that upon
construction of the neutrino factory the near detector sitting in front of
the storage ring will give stringent bounds on NSI, possibly 
even severer ones than currently imagined \cite{davidson}.  
Even in the case where the effects of NSI in three different places 
are comparable in size, it is unlikely that the feature obtained in 
our study with only propagation $\varepsilon$'s are completely 
cancelled by the effects of $\varepsilon$'s in production and 
detection processes.

As a theorist the natural question for me to ask is: 
``What would be the magnitude of $\varepsilon_{\alpha\beta}$?'' 
On dimensional ground the operator in (\ref{LNSI}) is suppressed 
by $M_{NP}^2$ \cite{concha1}. 
Since we normalize the operator with Fermi constant  $G_F$, 
$\varepsilon$ must be of the order of $(M_{W} / M_{NP} )^2 \sim$
0.01 (0.0001) if $M_{NP} = 1 (10)$ TeV.\footnote{
If we have to go to dimension eight operators their effective strength 
would be at most $(M_{W} / M_{NP} )^4 \sim 10^{-4}$ even for 
$M_{NP} = 1$ TeV. 
}
Therefore, the apparatus has to have sensitivity to the interactions with 
strength of 0.01\% $-$ 1\% of weak interactions to look for the effects of NSI. 
This is a highly demanding requirement.

\section{Which Apparatus?}

Let us consider which apparatus may be required to meet the 
condition of search for new interactions 100$-$10000 times weaker 
than weak interaction. 
To make a rough estimate let me assume, for brevity, that sensitivity to 
$\theta_{13}$ is comparable to that of $\varepsilon$. 
I expect, very roughly, that sensitivity to $\sin^2 2\theta_{13}$ is 
up to $\sim 0.01$ in conventional muon neutrino superbeam 
experiments \cite{superbeam}, 
which can be translated into $\varepsilon$ sensitivity of $\sim 0.05$.
Thus, most probably, superbeam is not the right apparatus as a 
machine to hunt NSI.
(We will however comments on its sensitivity later.)

As is well known, the alternative apparatus which is capable 
for looking into effects of smaller $\theta_{13}$ is either 
neutrino factory \cite{nufact} or beta beam \cite{beta}. 
Then, they are the good candidates for apparatus for hunting NSI. 
In my talk I concentrate on neutrino factory, 
leaving beta beam capability a subject of future studies by experts. 
For earlier analyses of NSI effects in neutrino factory, see e.g., 
\cite{gago,confusion1,confusion2,campanelli,ota1,kopp1}. 
We will see that the sensitivity to NSI by neutrino factory is fantastic.

\section{Problems in Neutrino Factory Search for NSI}

Unfortunately, it is known that one has to encounter inherent troubles 
in doing neutrino factory search for NSI. 
There exist two types of confusion problem:

\begin{itemize}

\item 
$\theta_{13}-$NSI confusion \cite{confusion1,confusion2}; 
The effects of non-vanishing $\theta_{13}$ can be mimicked by some 
of the NSI elements $\varepsilon$'s. 

\item
Two-phase confusion \cite{ota1}; 
The effects of leptonic Kobayashi-Maskawa (KM) phase 
$\delta$ can be imitated by the phases of the NSI elements 
$\varepsilon_{\alpha \beta}$, which will be denoted as $\phi_{\alpha \beta}$.

\end{itemize}

\noindent
The former confusion is fatal for precision $\theta_{13}$ 
measurement, while the latter one serious for identifying nature of CP 
violation even if it were observed.

\begin{figure}[h]
\vspace*{-0.2cm}
\begin{center}
\epsfig{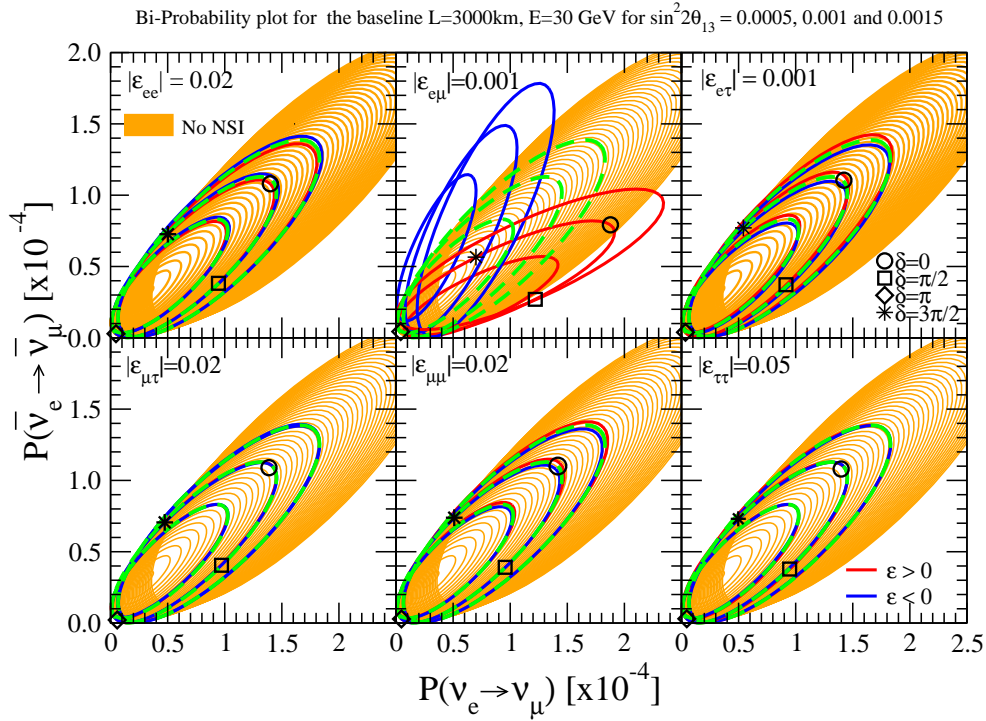}
\end{center}
\caption{
Bi-probability plots in 
$P(\nu_e \to \nu_{\mu}) - P(\bar{\nu}_e \to \bar{\nu}_{\mu})$ space 
at $L=3000$ km, for $E=$ 30 GeV, computed numerically 
using the constant matter density $\rho =  3.6$ g/cm$^3$ with 
the electron number density per nucleon equals to 0.5.  
The both axes is labeled in units of $10^{-4}$. 
In each panel only the indicated particular $\varepsilon_{\alpha \beta}$ 
is turned on. 
The upper (lower) panels, from left to right, correspond to the case
of non-vanishing $\varepsilon_{e e}$, $\varepsilon_{e \mu}$, and
$\varepsilon_{e \tau}$ ($\varepsilon_{\mu \tau}$, $\varepsilon_{\mu
\mu}$, $\varepsilon_{\tau \tau}$), respectively.  The red and the blue
ellipses are for positive and negative signs of $\varepsilon$,
respectively, for the cases with (from left to right) $\sin^2
2\theta_{13} = 0.0005$, 0.001, and 0.0015, as indicated in the
heading.
In the left and right lower panels the ellipses with positive and 
negative sign of $\varepsilon$ overlap almost completely and 
each individual curve is not visible. 
The green ellipses which correspond to the same three values of 
$\sin^2 2\theta_{13}$ but without NSI are clearly visible. 
}
\label{biP3000}
\end{figure}

It is not difficult to understand the causes of the two types of confusion. 
In Fig.~\ref{biP3000} presented are the bi-probability plots in 
$P(\nu_e \to \nu_{\mu}) - P(\bar{\nu}_e \to \bar{\nu}_{\mu})$ space 
\cite{MNjhep01}. 
The neutrino energy is taken to be $E=30$ GeV and the baseline 
$L=3000$ km. 
The blue and the red ellipses correspond to the case of positive and 
negative $\varepsilon_{\alpha \beta}$. 
Except for the case with $\varepsilon_{e \mu}$ these two are 
barely distinguishable. 
The orange ellipses are the bi-probability diagrams without NSI. 
There are so many of them because they are results of 
varying $\theta_{13}$. 
The point is that, apart from the case with $\varepsilon_{e \mu}$, 
the blue and the red ellipses are completely 
``absorbed'' into the background of orange ellipses. 
Namely, the system with NSI can be mimicked by adjusting 
$\theta_{13}$, the $\theta_{13}-$NSI confusion.


The two-phase confusion is also easy to understand. 
Let us ignore the solar $\Delta m^2_{21}$ assuming that 
it gives relatively small effect. 
The system is then reduced to an effective two generation problem. 
In such a system CP violating phase must be unique if a single 
type of off diagonal NSI element is introduced, because effects of 
the KM type phase must be (effectively) absent. 
Therefore, the two phases $\delta$ and $\phi_{\alpha \beta}$
must come together, the reasoning spelled out in \cite{NSI-nufact}. 
It was shown in pertubative computation \cite{ota1} that it is via 
the form $\delta + \phi_{\alpha \beta}$. 
This is nothing but the cause of the two-phase confusion.

\section{Two-Detector Setting in Neutrino Factory}

We ask questions: 
What is the way to look for effects of NSI  with highest possible sensitivities? 
What is the way to resolve the two confusion problems?
I argue that the two-detector setting, one at baseline $\sim$3000 km 
and the other at $\sim$7000 km, is the answer to these questions. 
It may be regarded as neutrino factory version of the two-detector setting 
discussed earlier \cite{MNplb97,T2KK}. 
Nonetheless, we will observe that the synergy between the two 
detectors in the present case is far more spectacular than the other cases.

You may ask ``why a detector at $\sim$7000 km?'' 
In the present context, there are two reasonings to motivate a 
far detector at $\sim$7000 km, which is sometimes called \cite{huber-winter} as 
the magic baseline, $\frac{a L}{4 E} = \pi$: 

\begin{itemize}

\item 
It was shown in a previous study \cite{mina-uchi} that the baseline 
comparable to the magic baseline gives the best sensitivity to 
measurement of the earth matter density. 
The relevant figure drawn by Uchinami-kun for his Mr. thesis 
is pasted in my previous Venice report \cite{Nutele07} as Fig.~1. 
(For a related work, see \cite{gandhi-winter}.) 
Measuring the matter density is equivalent to determine 
$\varepsilon_{e e}$ in our present language. 
Then, it is natural to suspect that a detector at the magic baseline 
can be a sensitive tool for detecting the effects of diagonal $\varepsilon$'s.

\item
The magic baseline is characterized as the baseline where 
the solar oscillation amplitude vanishes \cite{smirnov}, 
and hence the effect of CP phase $\delta$ is absent. 
Thanks to this property a detector at $L \sim$7000 km may be powerful 
in detecting effects of off-diagonal $\varepsilon$'s.

\end{itemize}

\begin{figure}[h]
\vspace*{-0.2cm}
\begin{center}
\epsfig{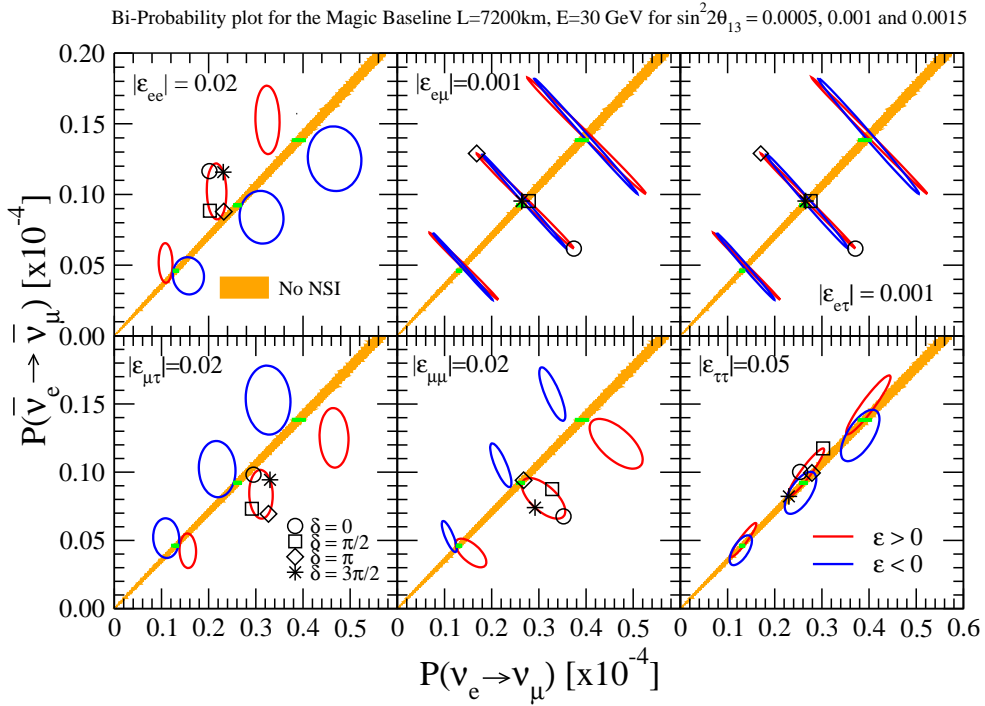}
\end{center}
\caption{
The same as in Fig.~\ref{biP3000} but for the baseline $L=7200$ km, 
the magic baseline, 
with the matter density $\rho = 4.5$ g/cm$^3$. 
The same values of $\varepsilon$ are used in each panel. 
}
\label{biP7000}
\end{figure}

\noindent
Because of the latter property it has been proposed \cite{intrinsic,huber-winter} that a second 
detector at the magic baseline is a powerful tool for resolving 
the conventional parameter degeneracy \cite{intrinsic,MNjhep01,octant}, 
in particular its intrinsic part. 
In fact, it allows us to have even higher sensitivity to off-diagonal 
$\varepsilon_{\alpha \beta}$. 
This is demonstrated in Fig.~\ref{biP7000}, in which the bi-probability 
plots in $P(\nu_e \to \nu_{\mu}) - P(\bar{\nu}_e \to \bar{\nu}_{\mu})$ 
space at $L=7200$ km are presented. 
As is clear in Fig.~\ref{biP7000} the ellipses without NSI shrink 
into points because of the absence of $\delta$ dependence, giving 
orange strips when $\theta_{13}$ is varied.  
On the other hand, the ellipses with NSI stand out. 
This property is nothing but the secret behind extremely high 
sensitivity to NSI which we will discover later.

In fact, we observe a prominent feature in systems with 
$\varepsilon_{e \mu}$ and $\varepsilon_{e \tau}$
that (1) the ellipses shrink to lines, and  
(2) they look identical.  
These features are easy to understand if one derives the approximate 
analytic formulas of oscillation probabilities. 
See \cite{NSI-nufact} for details. 
The one with $\varepsilon_{e \tau}$ is given as 
\begin{eqnarray}
&& P(\nu_e \to \nu_{\mu}; \varepsilon_{e \tau}) = 
4 \frac{ ( \Delta m^2_{31} )^2 }{  (  a - \Delta m^2_{31} )^2  } 
s^2_{23} s^2_{13} 
\sin^2 \left(\frac{\Delta m^2_{31} L}{4E}  \right)  
\nonumber \\
&+&
\hskip -0.2cm 
\frac{4 a  c_{23} s_{23}^2}{(a - \Delta m_{31}^2)^2}
  \Bigl[ 2  \Delta m_{31}^2 s_{13} \vert \varepsilon_{e\tau} \vert 
  \cos ( \delta + \phi_{e \tau} )
  + c_{23} a |\varepsilon_{e\tau}|^2 \Bigr] 
 \sin^2 \left( \frac{\Delta m_{31}^2 L}{4E} \right). 
\label{Pemu_magic-etau}
\end{eqnarray}
The corresponding formula for anti-neutrinos can be obtained by making
the replacement $a \rightarrow -a $, $\delta \rightarrow -\delta $,
and $ \phi_{e \tau} \rightarrow - \phi_{e \tau} $.
The formula with $\varepsilon_{e \mu}$ can be obtained by replacing 
$c_{23} \varepsilon_{e \tau} $ by $ s_{23} \varepsilon_{e \mu} $ 
in the second line of Eq.~(\ref{Pemu_magic-etau}), which explains 
the feature (2) above. 
The property (1), shrunk ellipse, is also evident by looking into 
(\ref{Pemu_magic-etau}); 
Since there is only  $ \cos ( \delta + \phi_{e \tau} ) $ dependence 
the ellipse must shrink into a line. 
Notice that at magic baseline the solar $\Delta m^2_{21}$ effect 
is absent and hence the two phase has to come together, 
as we have argued before and as indicated in 
(\ref{Pemu_magic-etau}).

\section{How Does the Two-Detector Setting Solve $\theta_{13}-$NSI Confusion?}

Before we discuss the sensitivity to NSI, let us first address 
the question of how the problem of $\theta_{13}-$NSI confusion
can be solved by the two detector setting at $L=3000$ and 7000 km. 
Unless we are able to solve this problem it is not practical 
to speak about neutrino factory as a hunting tool for NSI. 
It should be noticed that if NSI exists at the magnitude we anticipate in 
the present discussion and the effects of $\theta_{13}$ is comparable 
to that we inevitably have such the confusion. 
Therefore, this is not the problem only for neutrino factory, but for 
any other apparatuses which explore such region of mixing parameters.  
%

\begin{figure}[h]
\vspace*{-0.5cm}
\begin{center}
\epsfig{figure=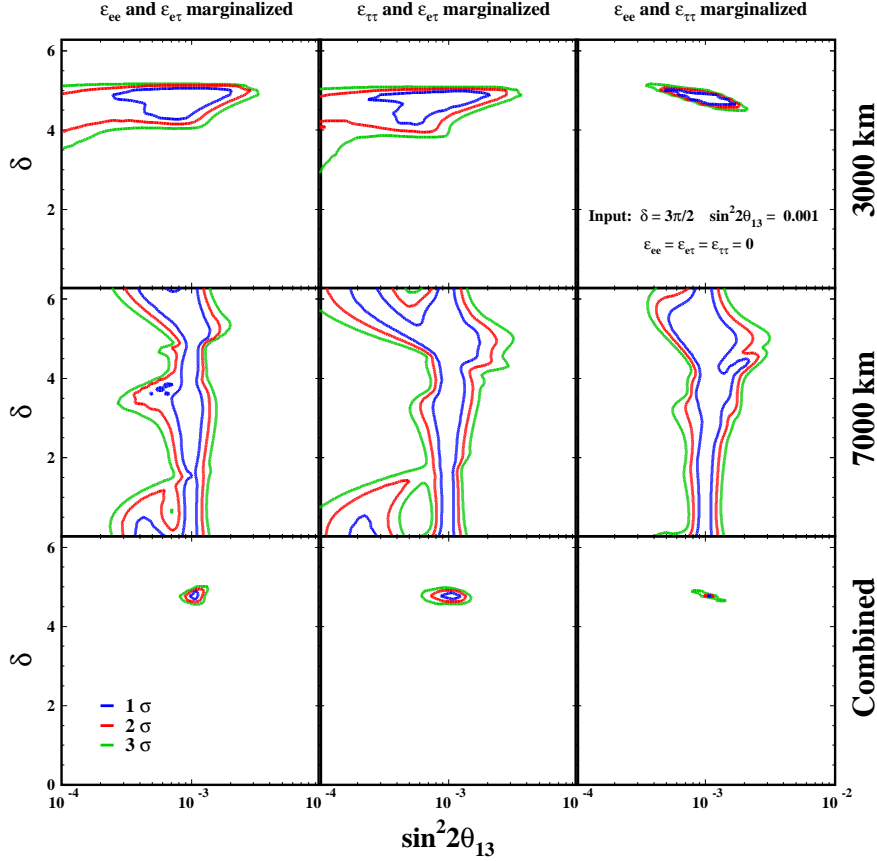,width=12cm}
\end{center}
\caption{
Allowed regions projected into the plane of 
$\sin^2 2\theta_{13}$-$\delta$ 
corresponding to the case where 
the input parameters are $\sin^2 2\theta_{13} = 0.001$ 
and $\delta = 3\pi/2$ and no non-standard interactions
(or all the $\varepsilon$'s are zero),  
for $E_\mu$ = 50 GeV and the baseline of $L=3000$ km (upper panels), 
7000 km (middle horizontal panels) and combination (lower panels). 
The fit was performed by varying freely 4 parameters, 
$\theta_{13}$, $\delta$ and 2 $\varepsilon$'s
where $\varepsilon_{ee}$ and $\varepsilon_{e\tau}$ 
are marginalized (left panels), 
$\varepsilon_{\tau\tau}$ and $\varepsilon_{e\tau}$ 
are marginalized (middle panels) 
and $\varepsilon_{ee}$ and $\varepsilon_{\tau\tau}$ 
are marginalized (right panels).  
}
\label{nsi-th13-del-3piby2-tau}
\end{figure}

\begin{figure}[h]
\vspace*{-0.5cm}
\begin{center}
\epsfig{figure=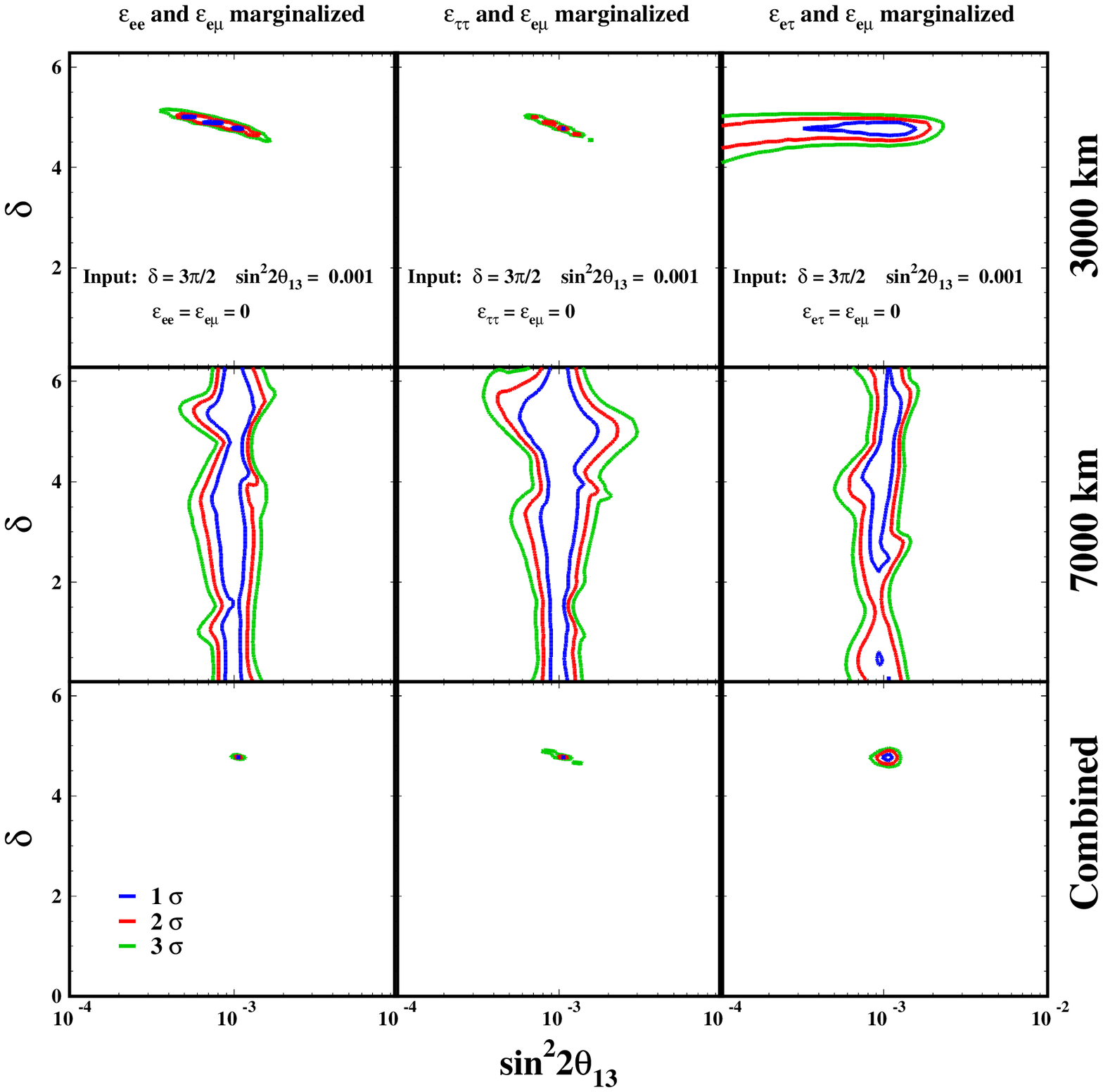,width=12cm}
\end{center}
\caption{
The same as in Fig.~\ref{nsi-th13-del-3piby2-tau}  but 
for different combination of 2 $\varepsilon$'s 
to which the fit to $\sin^2 2\theta_{13}$ and $\delta$ is marginalized;  
$\varepsilon_{ee}$-$\varepsilon_{e\mu}$ 
(left panels), $\varepsilon_{\tau\tau}$-$\varepsilon_{e\mu}$ (middle panels)
and $\varepsilon_{e\mu}$-$\varepsilon_{e\tau}$ (right panels).  
}
\label{nsi-th13-del-3piby2-mu}
\end{figure}

The results presented in this articles are based on \cite{NSI-nufact}. 
Therefore, the readers are advised to consult the reference 
whenever more detailed informations are necessary. 
In short our analysis assumed:
The number of muons decays per year is $10^{21}$, 
the exposure considered is 4 (4) years for neutrino (anti-neutrino),
and each detector mass is assumed to be 50 kton. 
The efficiency is assumed to be 100\% and the background 
is ignored.\footnote{
Alternatively, one may regard this setting as 5+5 years running with 80\%
efficiency, which may not be so far from the reality. 
}

In Fig.~\ref{nsi-th13-del-3piby2-tau} and 
Fig.~\ref{nsi-th13-del-3piby2-mu}, presented are the allowed regions 
projected into the plane of $\sin^2 2\theta_{13}$-$\delta$ 
corresponding to the cases with various combinations of NSI 
parameters which are turned on. The input parameters are taken as 
$\sin^2 2\theta_{13} = 0.001$, $\delta = 3\pi/2$, and 
$\varepsilon_{\alpha \beta}=0$. 
In the top panels (which show the constraint placed by the 
detector at $L=3000$km) 
the $\theta_{13}-$NSI confusion is clearly visible in most cases 
except for the panels involving $\varepsilon_{e\mu}$. 
Despite the vanishing input of NSI parameters, the freedom of 
adjusting them to nonvanishing values during the fit creates 
the $\theta_{13}-$NSI confusion. 
An exceptional situation occurs in the systems with $\varepsilon_{e\mu}$; 
The $\theta_{13}-$NSI confusion is much milder than that in other systems. 
This is, of course, expected from the behavior of ellipses in 
Fig.~\ref{biP3000}.

\begin{figure}[h]
\vspace*{-0.5cm}
\begin{center}
\epsfig{figure=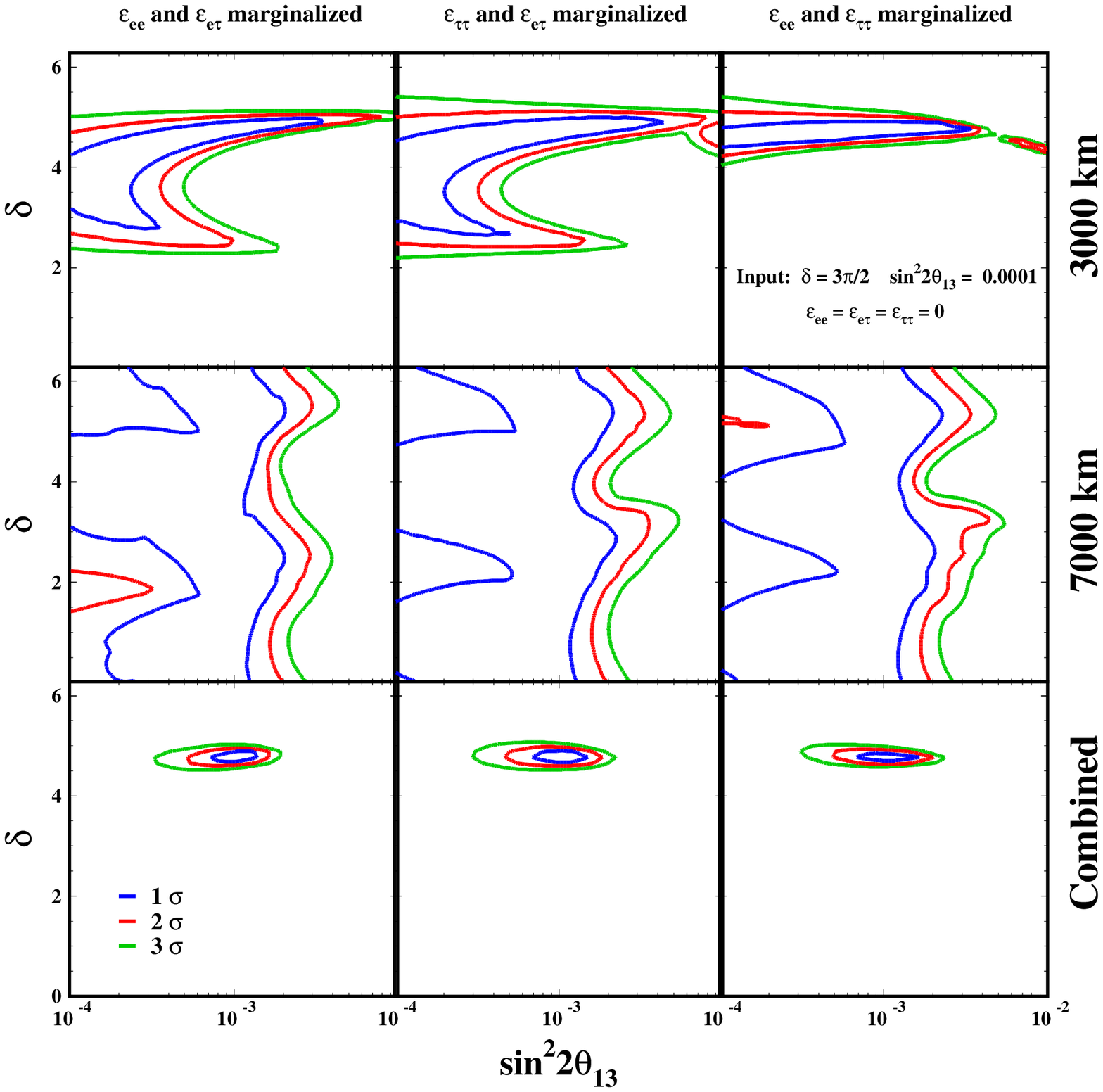,width=12cm}
\end{center}
\caption{
The same as in Fig.~\ref{nsi-th13-del-3piby2-tau}  but with 
$\sin^2 2\theta_{13}=0.0001$.
}
\label{nsi-th13-del-3piby2-tau-0.0001}
\end{figure}

We notice that the extent of the confusion depend on many things, 
e.g., on which combination of NSI parameters are turned on. 
In particular, the confusion is much severer for smaller $\theta_{13}$ 
as shown in Fig.~\ref{nsi-th13-del-3piby2-tau-0.0001} in which 
$\sin^2 2\theta_{13}=0.0001$. 
For the corresponding figure for the cases with $\varepsilon_{e\mu}$ 
and for dependence on  $\delta$, see Figs.~14 and 
Figs.~7-10, respectively,  in \cite{NSI-nufact}.

We observe in the bottom panels in 
Fig.~\ref{nsi-th13-del-3piby2-tau},  
Fig.~\ref{nsi-th13-del-3piby2-mu}, and 
Fig.~\ref{nsi-th13-del-3piby2-tau-0.0001} 
that the confusion is resolved by adding the informations 
gained by the detector at $L=7000$km which are shown 
in the middle panels. 
The far detector has little sensitivity to $\delta$, as expected, 
but it has a good sensitivity to $\theta_{13}$, and hence has 
potential of resolving the  $\theta_{13}-$NSI confusion. 
This is analogous to the role played by the far detector at the 
magic baseline which  helps resolving the conventional 
neutrino parameter degeneracy.

\section{Synergy of Two Detectors and Sensitivity to NSI}

\begin{figure}[h]
\vspace*{-0.5cm}
\begin{center}
\epsfig{figure=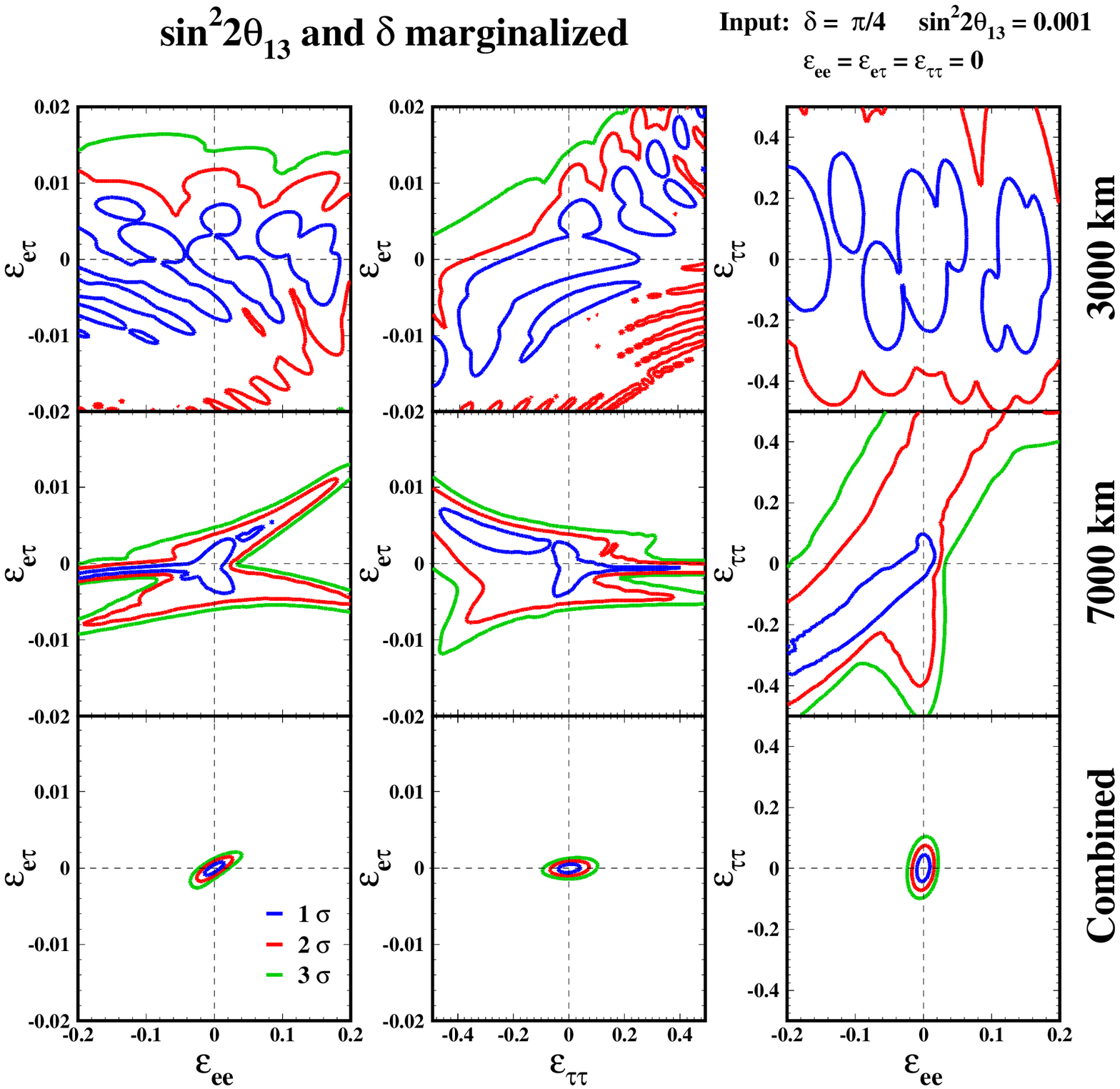,width=12cm} 
\end{center}
\caption{
Allowed regions projected into the plane of 
2 NSI parameters, $\varepsilon_{ee}$-$\varepsilon_{e\tau}$
(left panels), $\varepsilon_{\tau\tau}$-$\varepsilon_{e\tau}$ (middle panels)
and $\varepsilon_{ee}$-$\varepsilon_{\tau\tau}$ (right panels) 
corresponding to the case where 
the input parameters are $\sin^2 2\theta_{13} = 0.001$ 
and $\delta = \pi/4$ and no non-standard interactions
(or all the $\varepsilon$'s are zero), 
for $E_\mu$ = 50 GeV and 
the baseline of $L=3000$ km (upper panels), 7000 km (middle horizontal panels) 
and combination (lower panels). 
The thin dashed lines are to indicate the input values of 
$\varepsilon_{\alpha \beta}$.
The fit was performed by varying freely 4 parameters, 
$\theta_{13}$, $\delta$ and 2 $\varepsilon$'s 
with $\theta_{13}$ and $\delta$ being marginalized. 
}
\label{ee-et-tt-piby4}
\end{figure}

\begin{figure}[h]
\vspace*{-0.5cm}
\begin{center}
\epsfig{figure=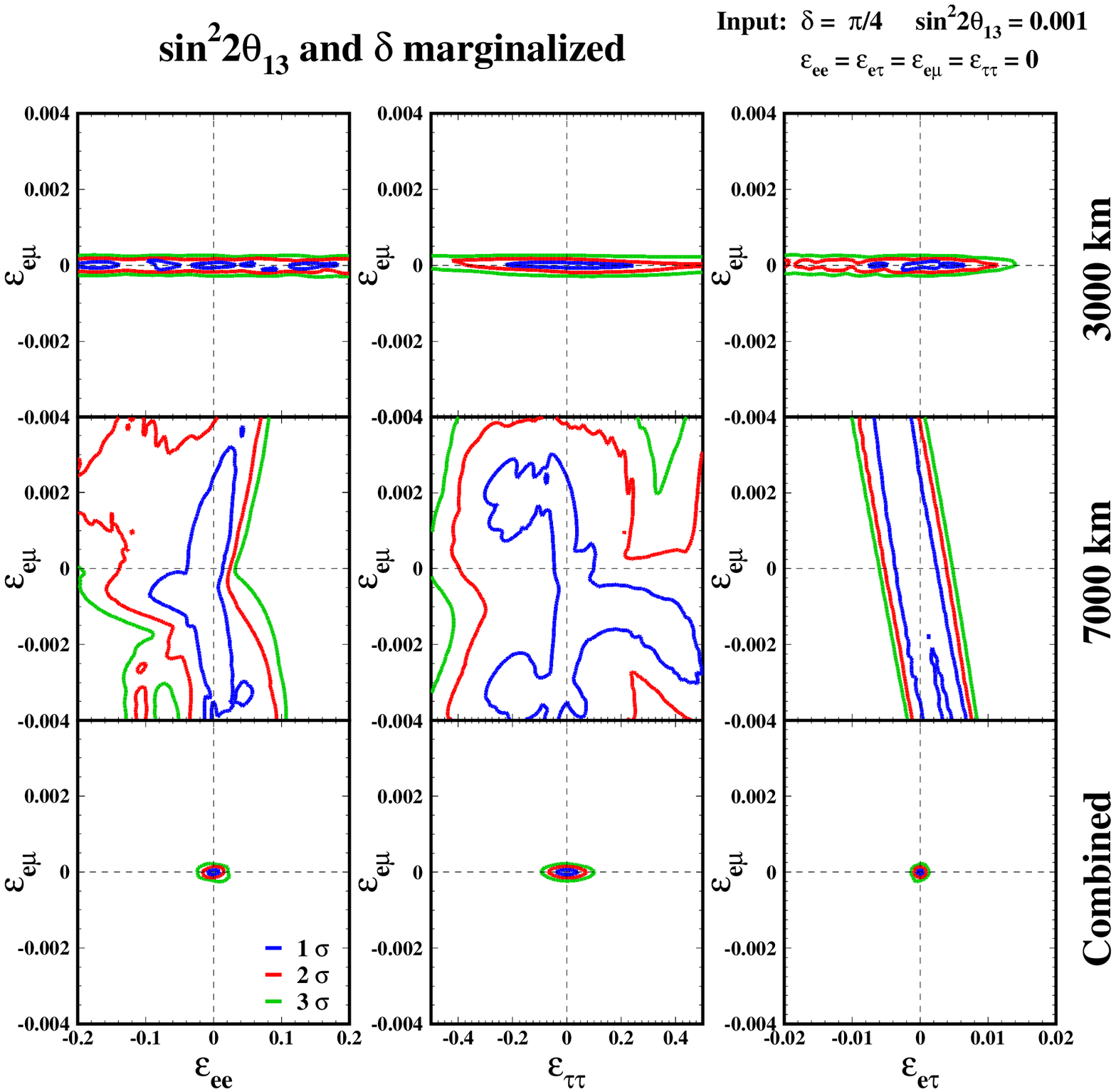,width=12cm}
\end{center}
\caption{
The same as in Fig.~\ref{ee-et-tt-piby4} but  
for a different combination of 2 $\varepsilon$'s, 
$\varepsilon_{ee}$-$\varepsilon_{e\mu}$
(left panels), $\varepsilon_{\tau\tau}$-$\varepsilon_{e\mu}$ (middle panels)
and $\varepsilon_{e\mu}$-$\varepsilon_{e\tau}$ (right panels). 
}
\label{ee-em-mm-piby4}
\end{figure}

Now, we turn to our original problem, the sensitivity to NSI 
possessed by the two-detector setting. 
The power of the synergy by the two-detector setting is enormous; 
Let us see it in Fig.~\ref{ee-et-tt-piby4} and Fig.~\ref{ee-em-mm-piby4}; 
Seeing is believing!

In Fig.~\ref{ee-et-tt-piby4} and Fig.~\ref{ee-em-mm-piby4}
presented are the allowed regions in space spanned by two of the 
NSI parameters $\varepsilon_{\alpha \beta}$ which are turned on 
in these particular simulations. 
The top, the middle, and the bottom panels are for the detector 
at $L=3000$ km, $L=7000$ km, and the two detector combined, 
respectively.

In Fig.~\ref{ee-et-tt-piby4}, we notice a remarkable synergy 
by the near (3000 km) and the far (7000 km) detectors. 
Normally, one does not expect that such a tiny allowed region 
emerges in the bottom panel by combing the ones in the top and the 
middle panels. 
The secret behind the extreme synergy is in the CP phase $\delta$; 
The region of apparent overlap between regions in the top and the 
middle panels differs in the fit value of $\delta$, and therefore 
disappear when two detectors are combined. 
It implies that keeping the solar $\Delta m^2_{21}$ is crucial to 
make the synergy active. 
Though it may sound trivial, I note that this effect is dropped 
off in many of the earlier treatment of NSI.

We have concluded as follows in our paper \cite{NSI-nufact}: 
``The sensitivities to off-diagonal $\varepsilon$'s are excellent, 
$|\varepsilon_{e \tau}| \simeq \mbox{a few} \times10^{-3}$ and 
$|\varepsilon_{e \mu}| \simeq \mbox{a few} \times10^{-4}$, 
while the ones for the diagonal $\varepsilon$'s are acceptable, 
$|\varepsilon_{e e}| (|\varepsilon_{\tau \tau}|) \simeq 0.1 (0.2)$ 
at 3$\sigma$ CL and 2 DOF. 
These sensitivities remain more or less independent of $\theta_{13}$ 
down to extremely small values such as $\sin^2 2\theta_{13}=10^{-4}$. 
They seem also very robust in the sense that they are 
not very disturbed by the presence of another non-zero NSI 
contribution.
The above characteristics of the sensitivities to NSI suggest that 
in our setting the off-diagonal $\varepsilon$'s are likely the best 
place to discover NSI.''
This last point was confirmed by a recent calculation \cite{kopp3}.

\section{Two-Phase Confusion}

Our treatment in \cite{NSI-nufact} does not contain full treatment 
of the two-phase confusion, but a partial one. 
We allowed negative values of $\varepsilon_{\alpha \beta}$, 
which can be interpreted as allowing two discrete values of phase 
$\phi_{\alpha \beta}=0$ and $\pi$. 
Therefore, we can in principle address the question of the 
two-phase confusion, its discrete version, in our treatment. 
%

\begin{figure}[h]
\vspace*{-0.5cm}
\begin{center}
\epsfig{figure=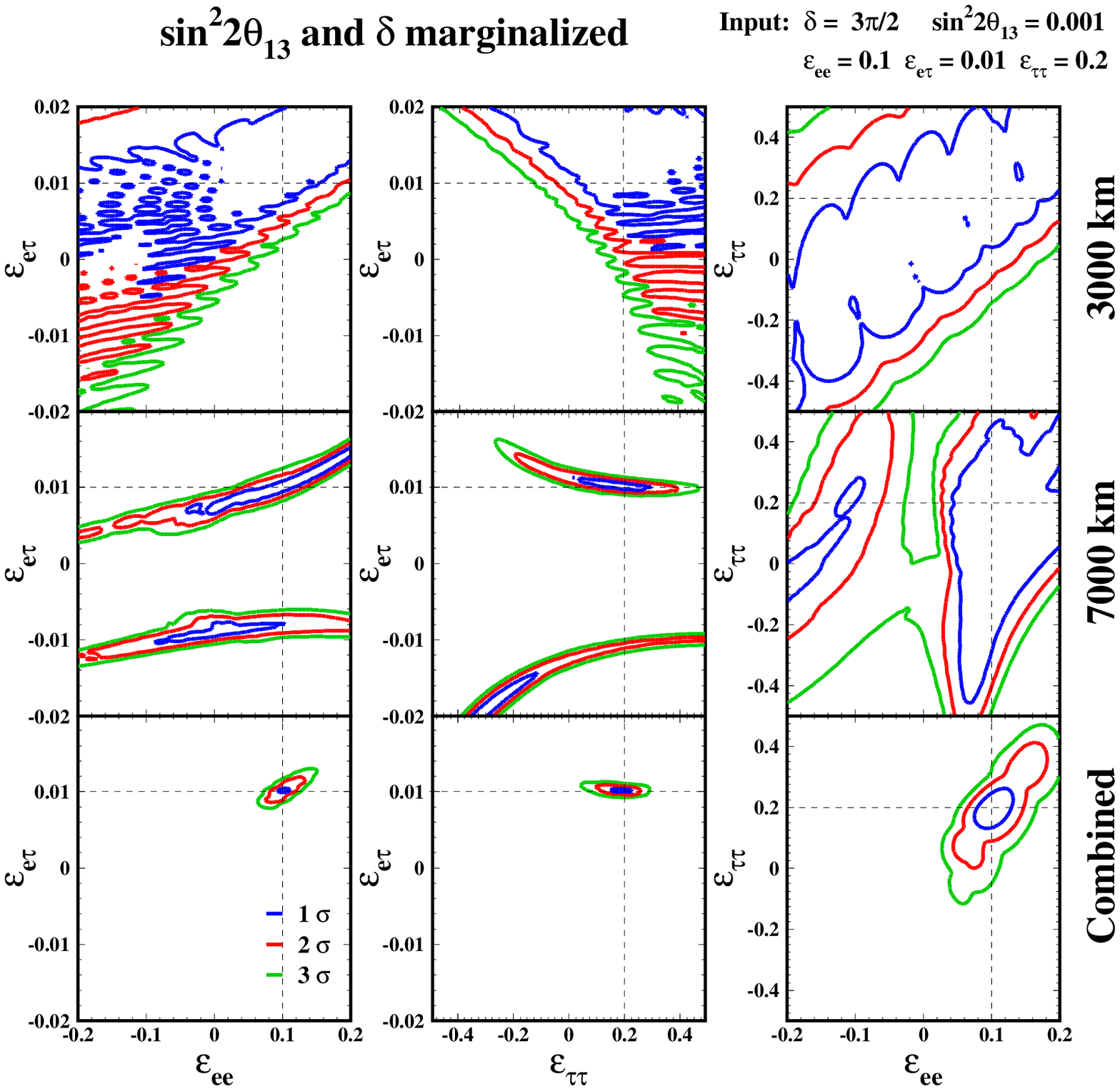,width=12cm}
\end{center}
\caption{
These figures are similar to those presented in Fig.~\ref{ee-et-tt-piby4} 
but for non-vanishing input values of $\varepsilon$; 
$\varepsilon_{ee} = 0.1$,  $\varepsilon_{e\tau} = 0.01$ and  
$\varepsilon_{\tau\tau} = 0.2$. 
We note that only the input values of 2 $\varepsilon$'s are 
set to be non-zero at the same time.  
The thin dashed lines indicate the corresponding non-zero values of 
$\varepsilon_{\alpha \beta}$ for each panel.
}
\label{ee-et-tt-3piby2-nzero}
\end{figure}

In Fig.~\ref{ee-et-tt-3piby2-nzero} we show the similar allowed regions 
but obtained in analysis with nonzero input values of NSI. 
In the middle panels in Fig.~\ref{ee-et-tt-3piby2-nzero}, which 
correspond to constraints imposed by the far detector,  
there are two discrete solutions of $\varepsilon_{e\tau}$. 
It is nothing but remnant of the two-phase confusion. 
Notice that there is no chance of resolving the confusion only by 
the detector at the magic baseline, as indicated in the expression 
of the oscillation probability in (\ref{Pemu_magic-etau}).

Again the synergy of the near and the far detectors makes it possible 
to resolve the discrete version of the two-phase confusion, 
as indicated in the bottom panels in Fig.~\ref{ee-et-tt-3piby2-nzero}. 
Though our treatment in \cite{NSI-nufact} did not allow us to fully 
address the issue, we expect that the two-phase confusion 
will be resolved by the two detector setting.

\section{Sensitivity to NSI by T2KK and the Related Settings}

So far we have confined ourselves into neutrino factory, and 
apparently there is little room for superbeam experiments as 
commented earlier. 
But, it is not completely true. 
As far as (2-3) (or $\mu-\tau$)  sector of the MNS matrix is concerned 
superbeam experiments with tuned beam energy to the one 
corresponding to the oscillation maximum is competitive to 
neutrino factory \cite{T2K,MSS04,resolve23}.

Therefore, I briefly discuss NSI sensitivity achievable by 
some of the superbeam experiments. 
For brevity I treat only three options with an upgraded beam 
of 4 MW from J-PARC: 

\begin{itemize}

\item

Kamioka-Korea setting:
Two identical detectors one at Kamioka and the other in Korea 
each 0.27 Mton fiducial mass 

\item
Kamioka-only setting: A single 0.54 Mton detector at Kamioka 

\item
Korea-only setting: A single 0.54 Mton detector at somewhere in Korea.

\end{itemize}

\noindent
The second option is nothing but the one described in LOI of T2K 
experiment as its second phase \cite{T2K}, which I call T2K II. 
The first one is sometimes dubbed as T2KK 
(abbreviation of Tokai-to-Kamioka-Korea),\footnote{
As I repeatedly emphasize, it is no more than a temporary name 
for idea of such apparatus. 
Even in the case people prefer one which succeeds to T2K, 
the last letter is naturally be the name of place (P if Pohang, for example) 
where Korean detector is placed. 
}
a modified version of T2K II by dividing the detector into 2 
and bring one of them to Korea \cite{T2KK}.

\begin{figure}[h]
\vspace*{-1.0cm}
\begin{center}
\epsfig{figure=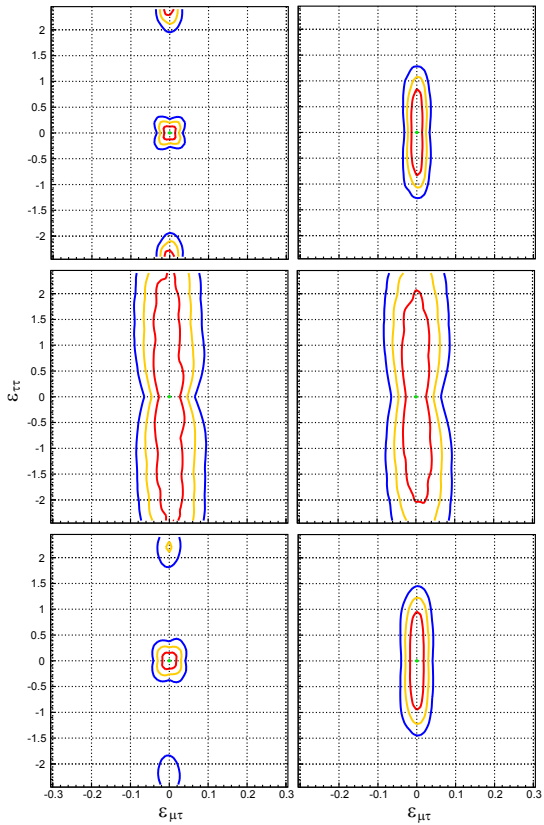,width=9cm}
\end{center}
\vspace*{-0.7cm}
\caption{
The allowed regions in
$\varepsilon_{\mu\tau} - \varepsilon_{\tau\tau}$ space for 4 years neutrino
and 4 years anti-neutrino running.
The upper, the middle, and the bottom three panels are for
the Kamioka-only setting, the Korea-only setting, and the
Kamioka-Korea setting,  respectively.
The left and the right panels are for cases with
$\sin^2 \theta \equiv \sin^2 \theta_{23} = 0.45$ and 0.5, respectively.
The red, the yellow, and the blue lines indicate the allowed regions
at 1$\sigma$, 2$\sigma$, and 3$\sigma$ CL, respectively, 
for 2 degrees of freedom. 
The input value of $\Delta m^2_{32}$ is taken as $2.5\times 10^{-3}$ eV$^2$. 
}
\label{sens-all}
\end{figure}

In Fig.~\ref{sens-all} presented are the sensitivities to NSI elements 
$\varepsilon_{\mu\tau}$ and $\varepsilon_{\tau\tau}$ 
achievable by, from top to bottom,  
T2K II, the Korea-only setting, and by T2KK. 
They are the results obtained by a truncated treatment of the 
$\mu-\tau$ sector done in \cite{NSP-T2KK}.
Though not spectacular the both T2K II and T2KK have reasonable 
sensitivities to NSI; 
The sensitivities of three experimental setups at 2 $\sigma$ CL 
can be read off from
Fig.~\ref{sens-all}. The approximate
2 $\sigma$ CL sensitivities of the
Kamioka-Korea setup  for
$\sin^2 \theta = 0.45 ~(\sin^2 \theta = 0.5)$ are:
\begin{equation}
| \varepsilon_{\mu\tau} |  < 0.03 ~(0.03),~~~
| \varepsilon_{\tau\tau} - \varepsilon_{\mu\mu}  | <  0.3 ~(1.2).
\end{equation}
Here, we neglected a barely allowed region near $|\varepsilon_{\tau\tau}| = 2.3$,
which is already excluded by the current data. 
The bound on $| \varepsilon_{\mu\tau} |$ above modestly improves 
the current bound obtained by analyzing atmospheric neutrino data of 
Super-Kamiokande and MACRO \cite{fornengo}.

The sensitivity to NSI by T2K II is slightly better than 
that of T2KK. I note, however, that if we examine wider class of 
new physics such as quantum decoherence, Lorentz violation, etc., 
the over-all performance of T2KK is the best among the above 
three settings, always remaining as the next best if not the best 
\cite{NSP-T2KK}.

\section{Bounds from Ongoing and Near Future Experiments}

It is a legitimate question to ask to what extent the ongoing 
and the near future experiments are powerful. 
Sensitivities to NSI by the MINOS experiments are examined in 
\cite{yasuda,friedland,blennow}. 
The sensitivities to $\varepsilon$ parameters are of order unity.
Possible contribution by OPERA experiment is also examined 
\cite{ota2,esteban,blennow2}
which however does not alter the situation. 
Combination of superbeam experiments with reactor is also 
considered \cite{kopp2} which entailed the sensitivities 
$ \varepsilon_{e \mu} \sim 0.2$ for NSI in propagation.

\section{Conclusion}

I have raised a question of whether a successful precision measurement 
of neutrino masses and the lepton mixing parameters is the 
{\em last} word for future neutrino experiments. 
As a possible candidate for ``the answer is No'' options, 
I examined the possibility that non-standard neutrino interactions 
outside the Standard Model 
can be uncovered by neutrino factory experiments. 
It, however, raises two serious issues, the $\theta_{13}-$NSI confusion 
and the two-phase confusion, 
which we proposed to be resolved by the near (3000 km) - 
far (7000 km) two detector setting. 
I would like to emphasize that the results obtained in our analysis 
is strongly indicative of the feature that neutrino factory can be used 
as a discovery machine for NSI while keeping its primary function 
of performing precision  measurement of the lepton mixing parameters.
I also touched upon the sensitivity to NSI search by some superbeam 
type experiments which utilizes neutrino beam from J-PARC.

\section{Acknowledgements}

It was my fifth visit to Venice (as a scientist), but it was the most 
memorable one for me for many reasons. 
In particular, it was the first chance for me to breathe air outside 
Japan after my disease. 
I deeply thank Milla for her invaluable kind help offered to me 
in transportation from/to the airport, curing my limited ability to walk, 
without which my participation would not be possible. 
I am grateful to all of my collaborators, 
Hiroshi Nunokawa, Takaaki Kajita, 
Renata Zukanovich Funchal, Nei Cipriano Ribeiro, Pyungwon Ko,  
Shoei Nakayama, and Shoichi Uchinami, for fruitful collaborations.  
I was benefited by conversations with Osamu Yasuda and Noriaki Kitazawa. 
This work was supported in part by KAKENHI, Grant-in-Aid for
  Scientific Research, No 19340062, Japan Society for the Promotion of
  Science.

\end{document}